\documentclass[conference]{IEEEtran}
\IEEEoverridecommandlockouts
\usepackage[noend]{algpseudocode}
\usepackage{amsmath,amssymb,amsfonts}
\usepackage[labelformat=simple]{subcaption}

\usepackage[hidelinks]{hyperref}
\usepackage[utf8]{inputenc}
\usepackage{dblfloatfix}
\usepackage{tcolorbox}
\usepackage{algorithm}
\usepackage{listings}
\usepackage{graphicx}
\usepackage{booktabs}
\usepackage{textcomp}
\usepackage{graphicx}
\usepackage{cleveref}
\usepackage{multirow}
\usepackage{textcomp}
\usepackage{balance}
\usepackage{comment}
\usepackage{siunitx}
\usepackage{diagbox}
\usepackage{xcolor}
\usepackage{pifont}
\usepackage{bm}
\usepackage{diagbox}
\usepackage{setspace}

\usepackage[
backend=biber
,style=ieee
,minbibnames=1
,maxbibnames=99
,maxcitenames=1
,mincitenames=1
,eprint=false
,doi=false
,isbn=false
,url=false
]{biblatex}
\bibliography{ref} 
\AtBeginBibliography{\footnotesize}

\def\BibTeX{{\rm B\kern-.05em{\sc i\kern-.025em b}\kern-.08em
    T\kern-.1667em\lower.7ex\hbox{E}\kern-.125emX}}

\usepackage{tikz}
\tikzset{%
pics/mycirc/.style args={#1}{
      code = {
\node [draw, circle,fill, scale=0.5] {\color{white}#1};
}}}

\begin{document}

\title{Towards a Complete Metamorphic Testing Pipeline}

\author{
\IEEEauthorblockN{
	Alejandra Duque-Torres, Dietmar Pfahl}
		
	\IEEEauthorblockA{%
	\textit{Institute of Computer Science}, \textit{University of Tartu}, Tartu, Estonia \\
	E-mail: \{duquet, dietmar.pfahl\}@ut.ee}
}

\maketitle

\begin{abstract}
Metamorphic Testing (MT) addresses the test oracle problem by examining the relationships between input-output pairs in consecutive executions of the System Under Test (SUT). These relations, known as Metamorphic Relations (MRs), specify the expected output changes resulting from specific input changes. However, achieving full automation in generating, selecting, and understanding MR violations poses challenges. Our research aims to develop methods and tools that assist testers in generating MRs, defining constraints, and providing explainability for MR outcomes. In the MR generation phase, we explore automated techniques that utilise a domain-specific language to generate and describe MRs. The MR constraint definition focuses on capturing the nuances of MR applicability by defining constraints. These constraints help identify the specific conditions under which MRs are expected to hold. The evaluation and validation involve conducting empirical studies to assess the effectiveness of the developed methods and validate their applicability in real-world regression testing scenarios. Through this research, we aim to advance the automation of MR generation, enhance the understanding of MR violations, and facilitate their effective application in regression testing. 
\end{abstract}


\section{Introduction}
\label{sec:introduction}

\textit{Metamorphic Testing} (MT) is a software testing approach proposed by \citeauthor{chen2020metamorphic}~\cite{chen2020metamorphic} to alleviate the test oracle problem. The test oracle problem arises when the SUT lacks an oracle or when developing one to verify the computed outputs is practically impossible \cite{DuqueTorres2020UsingRM}. Unlike traditional testing techniques, MT analyses the relations between pairs of input-output combinations across consecutive executions of the SUT rather than focusing solely on verifying individual input-output combinations. Such relations between SUT inputs and outputs are known as Metamorphic Relations (MRs). MRs specify how the outputs should vary in response to specific input changes. When an MR is violated for at least one valid test input, it indicates a high probability of a fault within the SUT. Nevertheless, the absence of MR violations does not guarantee a fault-free SUT.  However, its effectiveness heavily relies on the MRs employed. A good MR must not only specify correctly the input-output relations across the valid input data space - it also must be capable of detecting incorrect program behaviour resulting from a fault in the SUT code.

The process of generating appropriate MRs is indeed a complex task that requires a comprehensive understanding of the SUT and its application domain. As a result, the selection of MRs is predominantly performed manually, relying on the expertise and knowledge of the testers or developers involved. Nevertheless, several approaches have been proposed for automatic MR generation and selection. For instance, \citeauthor{PMR1}~\cite{PMR1,PMR2,PMR3,PMR4,PMR5} proposed machine learning-based (ML) methods for predicting the applicability of MRs to scientific programs by analysing their source code in the form of control flow graph (CFG). \citeauthor{10.1145/2642937.2642994}~\cite{10.1145/2642937.2642994} presented a search-based algorithm for inferring polynomial MRs in numerical programs based on the analysis of program inputs and outputs. \citeauthor{TROYA2018188}~\cite{TROYA2018188} proposed a tool-supported method for the automated generation of MRs in model transformation programs using patterns and execution trace analysis. \citeauthor{10.1145/3468264.3473920}~\cite{10.1145/3468264.3473920} introduced a multi-objective search algorithm for generating numerical MRs by iteratively modifying test assertions to minimise the number of false positives and false negatives compared to a set of correct and incorrect executions of the SUT.

While those approaches have achieved promising results for automatic MR generation, they tend to be domain-specific or rely on strong assumptions, \textit{i.e.,} that a chosen MR must always apply to the valid input data space in its entirety. Automatically discerning whether an MR violation is caused by a fault in the SUT or due to the MR's inability to satisfy a specific behaviour of the SUT for certain test data,\textit{ i.e.}, for subsets of the input data space for which the MR simply doesn't apply, is a significant challenge. In current practice, interpreting an MR violation is largely a manual effort and, therefore, time-consuming and resource-intensive. Additionally, the cost of MT is directly affected by the number of MRs utilised. As the number of MRs increases, the number of test cases may grow exponentially. Consequently, this leads to longer execution times and a greater need for manual inspection of MR violations \cite{prioritization1, 8539189}. 

\section{Problem \& Research Statements}
\label{prblem_researchStatements}
In the context of generating, selecting, and discerning the causes of MR violations, we acknowledge that achieving full automation is challenging for two main reasons. Firstly, MRs are inherently dependent on the specific behaviour and characteristics of the SUT and its application domain. This means that a deep understanding of the SUT and its context is required to identify relevant MRs. Additionally, some MRs may be straightforward and easily identifiable, while others may require complex reasoning or domain-specific knowledge. Secondly, MRs can exhibit different levels of applicability across the valid input data space. It is crucial to consider subsets of the input data space where an MR may not be applicable due to specific conditions or constraints. However, we strongly believe that the process of generating, selecting, and defining constraints to distinguish whether an MR violation is caused by a fault in the SUT or due to the MR's incompatibility with specific test data can be partially automated. This partial automation can be achieved by incorporating the tester as a valuable source of support and direct feedback.


Our research is divided into three parts: (1) MR generation, (2) MR constraint definition, and (3) evaluation and validation of MR effectiveness and usefulness. In the MR generation part, we will explore automated techniques for generating and describing MRs based on a domain-specific language. In the MR constraint definition part, we will investigate methods for defining constraints that capture the nuances and conditions under which an MR may not be applicable. Finally, in the Evaluation and Validation part, we will evaluate the effectiveness and efficiency of the developed methods and tools through empirical studies and validate their applicability in real-world regression testing scenarios. The specific research goals and related research questions for each part are outlined below.

\subsection{MR Generation}
\label{MRGeneration}

In this part of our research, the focus is on identifying and understanding the potential sources from which MRs could be extracted for a targeted SUT. The key question we aim to address is: Where do relevant, diverse, and effective MRs come from? By investigating different possible sources, such as system specifications, domain knowledge, or even the source code of the targeted SUT, we aim to identify and extract MRs that can capture the relevant behaviour of the SUT. Another important task of this part is to explore methods for describing MRs such that they become machine-readable, applicable to different SUTs, and transferable to different domains. We aim to develop a representation or a domain-specific language that is suitable for expressing such MRs and will enable the automated translation of MRs into test code, facilitating their practical implementation. Our research goal (RG$_{1}$) in this part, therefore, is to provide a method for generating MRs and a generalised representation, or domain-specific language, for describing the generated MRs. 
To achieve RG$_1$, we must answer the following research questions:

\begin{itemize}
    \item RQ$_{1.1}$: How to efficiently generate relevant, diverse, and effective MRs?
    \item RQ$_{1.2}$: How to best represent the generated MRs in a machine-readable format?
\end{itemize}

The method for generating and describing MRs should be applicable to many kinds of SUTs and facilitate automated test code generation. 
By analysing system specifications, requirements, and domain knowledge, relevant SUT behaviours and input transformations will be identified to create effective MRs. Furthermore, a representation format, or domain-specific language, will be designed based on existing proposals made by others with more specific goals \cite{9159060,9270318,10089522}. To be useful, the representation format should be flexible, adaptable, and capable of accommodating a wide range of SUTs. 

\subsection{MR Constrain Definition}
\label{MRselecConstrain}

A strong statement of our research is that a good MR should not only accurately specify the input-output relations across the valid input data space but also be capable of detecting incorrect program behaviour resulting from faults in the SUT code. We strongly believe that a chosen MR does not always apply to the valid input data space in its entirety.  Instead, an MR can still be applicable for specified subsets of the valid input data space. The notion that an MR must always apply to the entire input data space may limit the effectiveness of the MR. In real-world scenarios, it is common for different subsets of the input data space to exhibit distinct behaviours and characteristics. By identifying these subsets and setting constraints based on test input data, we can enhance the effectiveness of MRs. Additionally, providing explanations for MR violations is crucial to understanding whether the violation is due to a fault in the code or a constraint imposed by the MR itself. This includes identifying patterns, trends, and underlying causes of violations. By analysing the reasons behind MR violations, testers can identify common patterns, potential faults, or limitations in the system, which can then be addressed and improved upon.

Our research goal (RG$_{2}$) in this part is to develop a method for defining constraints on MRs based on test data and providing explainability for MR outcomes (violation or non-violation). By incorporating test data constraints into the MRs for a specific SUT, we aim to improve their overall effectiveness. This approach will enable testers to gain a better understanding of the system's behaviour by observing the MR violations and non-violations explainability across different input data subsets. With this understanding, testers will be able to design more comprehensive and targeted test cases. 

To achieve RG$_2$, we must answer the following research questions:

\begin{itemize}
    \item RQ$_{2.1}$: How to correctly define constraints on MRs based on test data?
    \item RQ$_{2.2}$: How to explain the reasons for the MT verdict when using a specific MR?
\end{itemize}

\subsection{Evaluation and Validation}
\label{subsec:MREvaluationValidation}
The main goal of the evaluation and validation part (RG$_{3}$) is to assess the effectiveness and efficiency of the developed methods and tools through empirical studies. This will involve comparing it with manual MR selection approaches and fully automated methods. The objective is to validate the method's ability to identify relevant constraints for the MRs generated and enhance fault detection capabilities. To achieve this goal, we focus on answering the following research question:

\begin{itemize}
\item RQ$_{3.1}$: How effective are the proposed methods in finding faults?
\item RQ$_{3.2}$: How well do the proposed methods perform compared to existing approaches?
\end{itemize}

\section{MR Generation}
\label{sec:MRGeneration}
Drawing inspiration from open bug repositories, \citeauthor{7811322} developed METWiki—an MR repository. METWiki gathers MRs from approximately 110 applications of MT spanning diverse domains. The MRs available in METWiki have been extracted from an extensive literature review on MT \cite{segura2016survey}. This review explored a wide range of publications to identify and collect MRs used in various real-world scenarios. 
Despite the authors' exceptional efforts in creating METWiki, there are certain limitations that need to be acknowledged. First, the lack of updates since its initial publication in 2016 raises concerns about the currency and relevance of the MRs contained in METWiki. Second, the lack of uniform descriptions for the MRs in METWiki. This inconsistency in describing the MRs not only poses challenges for users in understanding and comparing them but also hinders the generation of automatic machine-readable representations. Third, the absence of correlations between application domains. While the MRs are categorised into eight domains based on their application, the lack of connections between these domains can limit the broader understanding and cross-domain utilisation of MRs. 

\textbf{Ongoing Work.} Building on the idea of METWiki, we are in the process of creating an updated database of MRs. Our approach involves considering the existing MRs in METWiki while also focusing on gathering MRs from papers presented at top conferences and specialised workshops such as the MET workshop at ICSE conference. The creation of this new database primarily involves manual effort. To ensure the quality and effectiveness of the database, we are evaluating various strategies to establish a uniform structure for MR descriptions and cross-domain categorisation. This effort aims to minimise redundancy and enhance the usability of the database. Our goal is to create a valuable resource that not only incorporates the existing MRs from METWiki but also encompasses a wider range of MRs reported in top conferences and workshops. This updated database will serve as a valuable reference for researchers and practitioners in the field of Metamorphic Testing, facilitating the discovery and application of relevant MRs in various software testing contexts. By doing so, we aim to answer \textbf{RQ$_{1.1}$}.

\textbf{Planned Work.} Once our database is completed, we will proceed to design a domain-specific language that serves as a bridge between the MR database and a machine-readable format. To achieve this, we will leverage existing proposals made by others, as outlined in \cite{9159060,9270318,10089522}. These proposals offer valuable insights and guidelines for developing a representation format that meets the criteria of flexibility, adaptability, and compatibility with various SUTs. By doing so, we aim to answer \textbf{RQ$_{1.2}$}. In addition to the previously mentioned limitation from METWiki, another important consideration is where to extract new MRs from, aside from the existing sources such as METWiki. Furthermore, once a set of MRs is available, it is crucial to determine how to match them with the SUT.

\textbf{Preliminary achievements.} To explore alternative sources for extracting MRs, we explored \citeauthor{BLASI2021111041}'s work on ``MeMo" \cite{BLASI2021111041}, specifically focusing on the MR-Finder module. This module infers MRs by analysing sentences in Javadoc's comments that describe equivalent behaviours between different methods of the same class. MR-Finder consists of three components: \textit{i)} A predefined set of 10 words representing equivalence (S10W). \textit{ii)} A mechanism using Word Move Distance (WDM) to measure semantic similarity between sentences. \textit{iii)} A binary classifier to identify sentences indicating MRs. Our research aimed to improve the MR-Finder module. We reconstructed the module and utilised the original dataset provided by the authors of MeMo to replicate their reported results and establish a baseline for further experiments. We explored two strategies (STRTG) to enhance MR-Finder. In STRTG No.1, we expanded the S10W set by adding more equivalent words. In STRTG No.2, we introduced a second template sentence to the MR-Finder module while keeping the S10W set unchanged.  Through successful re-implementation of the MR-Finder module, we achieved comparable results using the original S10W set. Our findings indicate that expanding the initial set of equivalent words, as demonstrated by STRTG No.1, is likely to improve the performance of MR-Finder. For more detailed information about our study and findings, we refer the reader to our publication in \cite{profes2022}. 

To match a pre-defined set of MRs with the SUT, we explored the Predicting Metamorphic Relations (PMR) approach proposed by \citeauthor{PMR1} \cite{PMR1,PMR2,PMR3,PMR4}. The idea behind PMR is to create a model that predicts whether a specific MR can be used to test a method in a newly developed SUT. 
We conducted a replication study on the PMR \cite{PMR3} to assess its generalisability across multiple programming languages \cite{Saner2022}. We rebuilt the preprocessing and training pipeline, closely replicating the original study. Our results verified the reported findings and formed the basis for further experiments. We also explored the re-usability of the PMR model trained on Java methods. We evaluated its applicability to functionally identical methods implemented in Python and C++. While the PMR model performed well with Java methods, its prediction accuracy significantly decreased for Python and C++ methods. However, we observed that retraining the classifiers on CFGs specific to Python and C++ methods improved performance. Additionally, we conducted an evaluation of the PMR approach using source code metrics as an alternative to CFG for building the models \cite{vst2022Aleja}. For more detailed information about our replication study and extension using source code metrics, we refer the reader to our publications in \cite{Saner2022} and \cite{vst2022Aleja}. 

\section{MR Constrain Definition}
\label{sec:MRConstrainDefinition}

Despite promising results from the original PMR study and subsequent works, the PMR has significant limitations. Firstly, it relies on binary classifiers that require labelled datasets to provide examples for learning. Labelled datasets may not always be available, and obtaining them can be time-consuming. Secondly, the feature extraction process for model training is based on CFG or source code metrics, which may not account for refactoring. This limitation can affect the accuracy of the PMR, as refactoring can change the structure of the code and, consequently, the way MRs apply. Lastly, the binary output of PMR may not consider test data and its impact on MR applicability. This limitation in PMR, where it does not consider scenarios with varying applicability of MRs to different test data with specific characteristics, can be referred to as constraints. Constraints are the conditions or limitations under which an MR may or may not be applicable. 

\textbf{Ongoing Work.} Motivated by the limitations of PMR and the challenges in selecting appropriate MRs, we are developing a novel approach called MetaTrimmer \cite{SANER-VST2023}, a test data-driven method for constraining MRs. Similar to PMR, we assume a pre-defined list of MRs is available. However, MetaTrimmer does not rely on labelled datasets and takes into account that an MR may only be applicable to test data with specific characteristics.  MetaTrimmer comprises three main steps: (1) Test Data Generation (TD Generation), (2) MT Process, and (3) MR Analysis. Step 1, TD Generation, is responsible for generating random test data for the SUT. In step 2, the MT Process carries out necessary test data transformations based on the MRs, and generates logs to record information about inputs, outputs, and any MR violations during the execution of the test data and the transformed test data against the SUT. MR Analysis, step 3, conducting manual inspections of violation and non-violation results and identifying specific test data or ranges where the MR is applicable to derive constraints. Moreover, we are developing a tool to support the MR Analysis step called MetaExploreX. By formalising and evaluating MetaTrimmer, we aim to answer RQ$_{2.1}$.

\textbf{Planned work.} In the next stage of our research, we plan to focus on automating the derivation of constraints. Currently, the process of identifying constraints involves manual inspection of the violation status, which can be a tedious and time-consuming task. By automating this process, we aim to increase the coverage of analysis and uncover potential constraints that may have been overlooked in manual inspections. This will enable testers to gain a better understanding of the constraints and applicability of MRs, allowing for more efficient and effective testing practices. Through data mining techniques, we will analyse the data associated with MR violations, test cases, and other relevant information to identify common patterns and associations. These findings will help us derive constraints that capture the specific conditions under which MRs are applicable or non-applicable. By doing so, we aim to answer RQ$_{2.2}$

\textbf{Preliminary achievements.}
We have introduced the main idea of MetaTrimmer and evaluated its effectiveness through a toy example. For more detailed information about this study, we refer the reader to our publication \cite{SANER-VST2023}. Furthermore, we have submitted and registered a paper in which we present the formalisation of MetaTrimmer and its evaluation on 25 Python methods and six-predefined MRs. A replication package with the full set of data generated during our experiments as well as all scripts 
can be found in our GitHub repo\footnote{\href{https://tinyurl.com/MetaTrimmer}{https://tinyurl.com/MetaTrimmer}}. The preliminary results obtained from this paper demonstrate a promising potential for MetaTrimmer. In addition to MetaTrimmer, we have developed the first prototype of MetaExploreX\footnote{\href{https://github.com/aduquet/MetaExplorex-docker}{https://github.com/aduquet/MetaExplorex-docker}}, a tool that offers visualisation and exploration capabilities for supporting the MR Analysis step of MetaTrimmer. 
\label{MREvaluationValidation}

\section{ Evaluation and Validation}
The evaluation and validation phase (RG$_{3}$) of our research aims to assess the effectiveness and efficiency of the developed methods and tools through empirical studies. One of the key objectives is to compare our approach with manual MR selection approaches and fully automated methods to validate its ability to identify relevant constraints for the generated MRs and enhance fault detection capabilities. To address the research question regarding the effectiveness of our approach (RQ$_{3.1}$), we plan to conduct a mutation testing analysis. Mutation testing involves introducing small modifications or ``mutants" into the SUT and examining how well the constrained MRs can detect these mutations as faults. 

In order to address research question RQ$_{3.2}$, which focuses on comparing the performance of our proposed methods to existing approaches, we will conduct a comprehensive comparison at every step of our research. Throughout the development and evaluation of our methods and tools, we will consistently compare their performance against baseline approaches. These baseline approaches may include existing manual MR selection methods or other fully automated approaches commonly used in the field.

\section{Final remarks }
\label{sec:ConcludingRemarks}
The effectiveness of MT has been demonstrated in various application domains, including autonomous driving~\cite{zhang2018deeproad,zhou2019metamorphic}, cloud and networking systems~\cite{canizares2020mt,9477667}, bio-informatics software~\cite{10.1145/3193977.3193981,shahri2019metamorphic}, scientific software~\cite{peng2021contextual,8533366}. However, achieving full automation in MT poses challenges due to the inherent dependence of MRs on the specific behaviour and characteristics of the SUT and its application domain.

In our research, we aim to provide assistance in the generation and constraint definition of MRs. By developing methods and tools that support testers in these processes, we aim to contribute to the advancement of the MT field. Our goal is to improve the automation and effectiveness of MR generation, taking into account the specific characteristics of the SUT and its application domain. We believe that the results of our research will produce concrete support for software testers, offering insights into effective strategies for MR generation and constraint definition. By doing so, we aspire to contribute to the advancement of MT as a reliable and valuable testing approach, guiding future research studies and fostering continuous improvement in this field.

\section*{Acknowledgement}
The research reported in this paper has been partly funded by BMK, BMAW, and the State of Upper Austria in the frame of the SCCH competence center INTEGRATE [(FFG grant no. 892418)] part of the FFG COMET Competence Centers for Excellent Technologies Programme, as well as by the European Regional Development Fund, and grant PRG1226 of the Estonian Research Council.

\newpage

\balance
\printbibliography

\end{document}